# Transitioning Between Audience and Performer: Co-Designing Interactive Music Performances with Children

**Alina Striner**

Human-Computer Interaction Lab

iSchool | University of Maryland

College Park, MD, USA

algol001@umd.edu

**Brenna McNally**

Human-Computer Interaction Lab

iSchool | University of Maryland

College Park, MD, USA

bmcnally@umd.edu

## Abstract

Live interactions have the potential to meaningfully engage audiences during musical performances, and modern technologies promise unique ways to facilitate these interactions. This work presents findings from three co-design sessions with children that investigated how audiences might want to interact with live music performances, including design considerations and opportunities. Findings from these sessions also formed a *Spectrum of Audience Interactivity* in live musical performances, outlining ways to encourage interactivity in music performances from the child perspective.

## Author Keywords

Performance interaction; Sound and music computing; Participatory Design; Children; Media arts

## ACM Classification Keywords

H.5.m. Information interfaces and presentation (e.g., HCI): Miscellaneous

## Introduction

In March 2015, a train station in France unexpectedly became a collaborative performance space when an traveler playing a public piano was joined by an onlooker (Figure 1). With the transition of an audience member into an active participant in the composition, a simple melody developed into a rich harmony. This impromptu piano duet became a worldwide phenomenon within days, inspiring more than 21.5 million views to date [9].

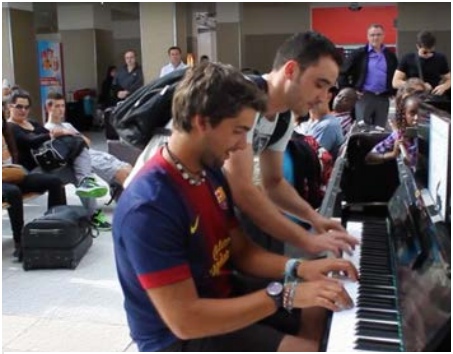

Figure 1: Strangers playing an impromptu duet at a train station in France [9]. The video has received over 21.5 million views on YouTube.

Music performance can be a difficult space for audience participation. While audiences can readily enjoy music, an understanding of structure is considered [1][12] fundamental to thoughtful contribution. Because of this, audience participation is often incorporated into performances for novelty or ornamentation [10]. However, modern tangible and multisensory technologies have the potential to facilitate new forms of audience participation in musical performances that overcome knowledge gaps and traditional barriers [15].

Previous works in this area have considered how novice audiences can participate in components of music creation; for instance, Hyperscore [7], a graphical sketchpad for novice composers, helped bridge the music notation knowledge gap, and music video games like Guitar Hero helped bridge the skill gap [1]. Since participation in music is positively linked to cognitive performance and self-esteem [21], it is important to support children to make sense of and appreciate complex music arts [20]. Today, we have the opportunity to accomplish this through expanding to different types of performance interactions. Rather than focusing on music making, this work expands upon previous ideas by investigating how children understand and wish to interact with different types of live music performances.

This exploratory work investigates the ways in which children can interact with live music performances as audience members through a series of three design sessions with children. Findings from these sessions describe children's expectations and perceived opportunities for audience participation in live music performance, and culminate in a *Spectrum of Audience Interactivity* for child audiences.

### Related Work
Here we review 1) types of audience participation in live musical performances and 2) technologies that facilitate audience participation in music performances.

*Types of Audience Participation in Musical Performances*
In music genres such as musical theater or opera, audience members and performers are segregated by stage, scaffolding, and lighting. Audiences sit in darkened silence, serenaded by performers who never see them. Performers are largely unable to gauge and react to audience reactions, as feedback such as clapping, talking, or cheering is often curbed to pre- and-post performance [16]. Unlike such traditional performance mediums, lighting in participatory mediums is relaxed and feedback is encouraged to add the "sense of liveness" [24]. As in the introductory example, street performances often encourage audiences to add to the performance [27]. Similarly, in gospel, call-and-response between performers and audiences nudges democratic participation [20], and in music improv, audience members may generate feedback through dance and other movements [21]. Although informal genres are more interactive, audience-performer interaction is often asynchronous or inequitable [16], giving higher degrees of interaction to audience members' closers to the stage [13] or in positions of power [19]. In this work, we explore how children envision their participation across traditional and participatory music performance spaces.

*Technologies that Facilitate Audience Participation*
Research in music and HCI has experimented with using technologies to support and augment performance interaction. For instance, Feitsch [8] digitally combined audience faces with a singer's, an opera allowed remote audiences to interact with an LED chandelier [14], and Freeman used an audience feedback loop to influence an orchestra [10]. Composer Eric Whitacre went beyond interaction and gave his YouTube audiences a voice; he created the Virtual Choir [29], a remote experience allowing singers to record and sync their voices around the world. Researchers have also begun to consider how to design for interactive performance experience. For instance, Maynes-Aminzade [18] developed a set of system and game design factors for interactive audiences, and Reeves [24,25] outlined interactions for crowd participation. Further, Rust's [26] work on performance experience design considers child storytelling co-creation in live performances. This work adds to these discussions, and considers how child audiences may use tangible and multisensory technologies to interact with live music performances. Although researchers have considered stages of audience interactions for music [2,17], to our knowledge no formal models of audience participation exist.

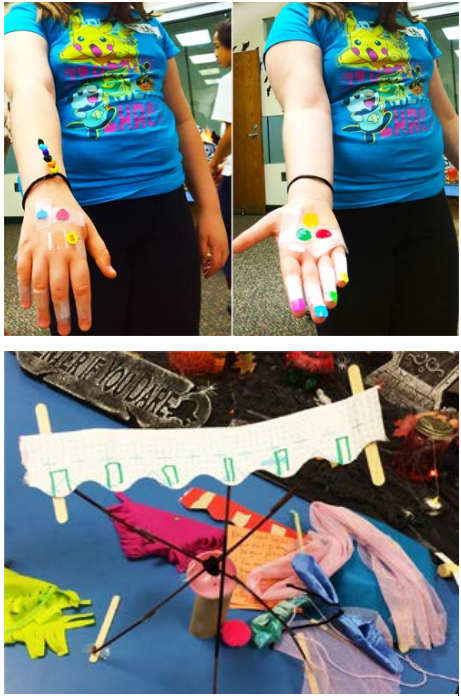

Figure 2: Designs from Session 1. At the top is a wearable set of buttons that allow audience members to make decisions during the performance, and below is a hologram keyboard that allows audience members to play along with the performance.

## Methods

To investigate how an audience might influence a live musical performance, three formative Cooperative Inquiry (CI) design sessions were held. CI is derived from Participatory Design, and brings together 6-8 children, ages 7-11, and interdisciplinary adults as equitable design partners who work together to design technology for children [5, 11]. CI has evolved to deliberately inform the design processes of children's technologies, and the design techniques used have been specifically modified to meet the needs of intergenerational teams [11]. Ten children in total participated in this work; not all children participated in all of the sessions. The three sessions investigated:

1. Interactions with A Piano Performer
2. Interactions with Any Performer
3. Varying Degrees of Interactivity

*CI Session Structure*. CI sessions all followed a similar structure [5,6,11]. Sessions began with a discussion of the day's design topic and goals. The team then divided into 3 small groups, each comprised of 2-3 children and 1-2 adults, and worked in these small intergenerational groups to design a technology that addressed the session's goals. At the end of each session, small groups presented their technology design ideas to the entire design team while an adult design partner wrote the Big Ideas on a whiteboard and performed a rapid thematic analysis [5]. The entire team then discussed and refined the themes and unique ideas that arose.

*Data and Analysis.* Across the three design sessions, at least one adult in each small group took observational notes. Session data included these notes, photographed session artifacts, audio recorded debriefs, and video recordings of presentations. Session themes (Big Ideas) were iterated through reviews of these data during discussions with the research team. The most prevalent themes from each design session are described in the session outcomes.

## SESSION 1: Interactions with a Pianist

To provide initial scope, in this session the team was asked to consider a performance that included a single performer and instrument: *How can an audience change what is being played during a piano performance?* Small groups used a 3D prototyping technique called Bags of Stuff [5] to build low fidelity prototypes of technologies that would interact with a live piano performance using art supplies.

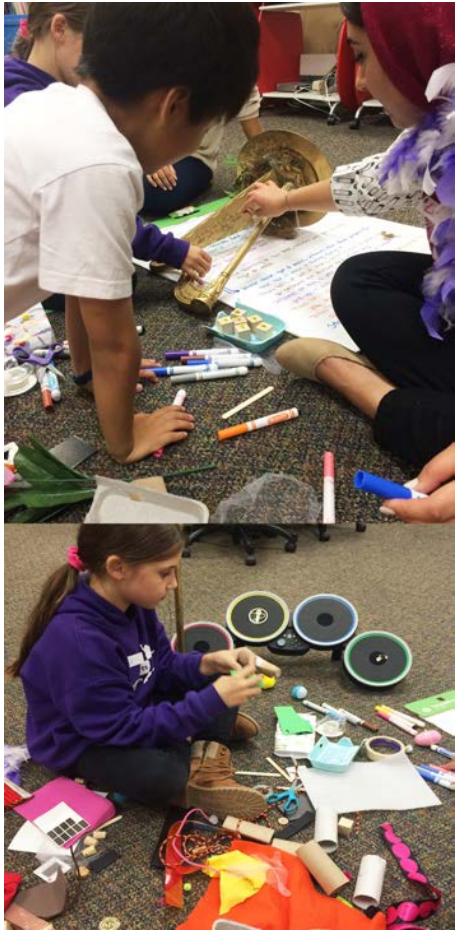

Figure 3: Design process during session 2. This session incorporated the Bags of Stuff technique with large musical stage props.

*Session 1 Outcomes*
*Tangible Experiences.* Groups primarily designed tangible, largely wearable, technologies such as interactive hats, hand sensors, and palm pushbuttons that could interface with the musical performance (Figure 2). Groups transformed both music and feedback into experiences that could be manipulated with tangible technologies. For instance, one group suggested creating tangible "sound chips," discrete bits of music that audiences could append to performance melodies. Another group suggested using a wearable forehead sensor to gauge audience engagement from head nodding and shaking.

*Addressing Personal and Group Preferences.* Groups made it clear that individual preferences for the experience of the performance must be considered, and designed technologies that encouraged personalized experiences. For instance, one group designed hats and earbuds that allowed them to add accompaniment to the piano performance that only they could hear. When making changes to the performance as a whole, aggregate feedback was generally sought to ensure everybody's desires would be addressed. For instance, children designed a button that changed the music based on audience members' overall mood.

## SESSION 2: Interactions with Any Performer

As with the previous session, the design team was asked to envision ways an audience could interact with a performance as it is happening. However, in this design session no constraints were placed on the type of performance being attended. This session used the Big Props technique [28], which incorporates large stage props (in this case, toy guitars, harps, etc.) in addition to art supplies to develop prototypes that focus on interactions (Figure 3). Small groups were instructed to design as many ways for audiences to interact with music performances as they could.

*Session 2 Outcomes*
*Multimodal Interactions.* In study 2, groups created multimodal interactions with which to engage in the performance. For instance, groups proposed using hand gestures to change sound effects, throwing tokens onstage to get better seats, clapping physical blocks to control volume, and waving LED pom-poms to conduct the musicians. Groups also suggested using tactile feedback, throwing colored paintballs at the stage, or using paper planes to send messages to each other.

*Multisensory and 4D Experiences.* All groups designed multisensory experiences into performances. For instance, one group wanted to control wind gusts to lift a bride's veil, another wanted to tickle a performer with remote controlled feathers, and a third wanted to graffiti a performer's outfit with a cellphone. Groups also wanted to influence their own multisensory experience; one group created a game played with performers where they could fall into ice water, and another had performers influence the type of food audience members had available during the show.

## SESSION 3: Varying Degrees of Interactivity

In design sessions 1 and 2 we observed that the design teams proposed a range of *individual to group* and *passive to active* interactions with performers. Inspired by these findings, the third design session began by asking small groups to review their designs to develop a spectrum of audience participation. Each group was given a list of previous ideas and asked to arrange them in order of least to most interactive (Figure 4). In the second half of the session, children individually voted on their favorite type of participation from the spectrum, and the team worked together to develop interactions to support the *becoming performers* category, which received the most votes (Figure 5).

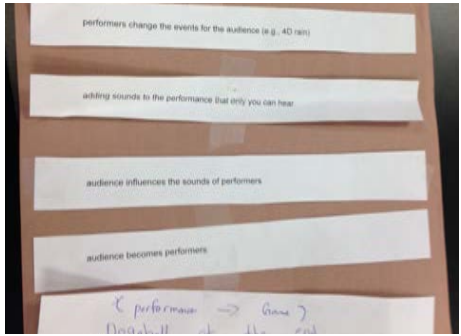

Figure 4: One group's ordering of interactivity

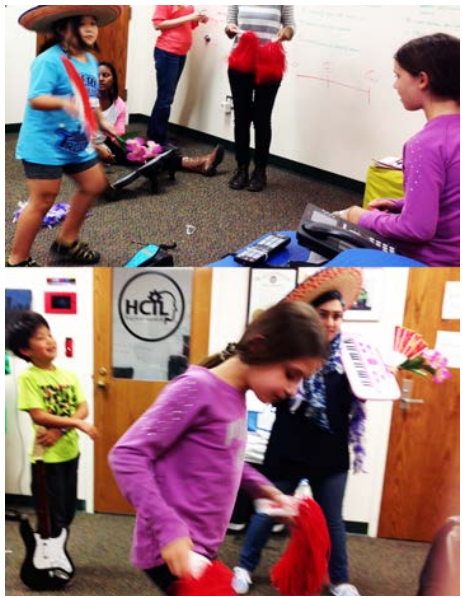

Figure 5: The entire team developing interactions for the *becoming performers* category. Audience members took turns watching and being on stage.

*Session 3 Outcomes*
*Rules for Democratization and Fairness.* The team strongly considered issues of fairness and democratization in their designs for interactive participation--either offering everyone a chance to participate, or including everyone and allowing individuals to opt out. In the democratized experience, the group proposed using a video camera that allowed audience members to play air guitar with the onstage performers from their seat. Similarly, in the fair experience, children sat in a line and took turns going onstage. While it was recognized that this dynamic created peer pressure on audience members, the team suggested this might keep audiences more engaged because they would be *"constantly on [their] toes."*

*Distinct Spaces.* Throughout the many designs there were clear boundaries between the audience and performance spaces. Audiences were either interacting from their seat, or distinctly transitioning from audience member to performer by going on stage. For instance, the group suggested that audience members could find out that they'd been selected by having their seat buzz, then go to a different room to become the performer, or could disappear into the group and reappear on stage. Interestingly, the group also emphasized a boundary between their role as an audience member or a performer; rather than participating throughout the performance, groups wanted audience members to enjoy the performance passively at times.

**A Spectrum of Audience Interactivity**
Findings from these CI sessions informed a *Spectrum of Audience Interactivity* in musical performances (Figure 6). The children's arrangements of ideas in session 3 were used to develop an initial spectrum (Figure 4). Session data on children's designs (e.g., artifact photos, notes, Big Ideas) were coded into this spectrum, resulting in refined level descriptions and the addition of a new level that distinguished between influencing and augmenting performances. The final spectrum represents a preliminary view into how children envision audience participation in live music performances on a scale of increasing interactivity.

The top of the spectrum indicates the lowest form of interactivity, 1) *passive observation*, which is common to traditional performance genres. More interactive is 2) *personalization of experience,* where audience members customize the performance for themselves; while this can be highly interactive, it is a private, rather than group, experience. In 3) *reaction to performance*, audience members react to the performance through visual, audio, or physical mediums (e.g., throwing tomatoes). By 4) *influencing events*, audience members impact decisions that happen on stage, such as what genre of music is played or what instruments performers use. Related to 4, in 5) *augmenting performance,* audience members directly add to a performance from their seats. For instance, audience members could hold up a lighter during a song. In 6) *augmenting the audience's multisensory experience*, performers break the "forth wall" by reacting to audience feedback, influencing the audience's space using multisensory elements like wind, rain, or selection of food. Finally, in 7) *becoming performers*, audience members take on comparable roles to performers, or take over the performance altogether. For instance, an audience member might join performers in a drum circle.

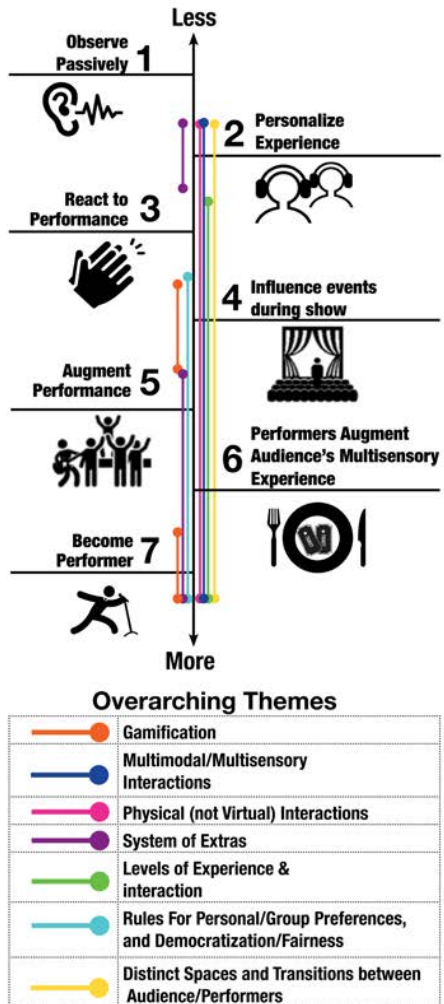

Figure 6. A spectrum of audience interactivity and an overview of how the themes presented across different levels of the spectrum.

## Discussion

The preliminary findings and spectrum presented in this work are a first step toward understanding the many ways in which technology can allow audience members to interact with performance in traditional and new music forms. While we focused on children, the prevalence of viral videos featuring audiences interacting with performance [13] suggests that adults may also be interested in interacting with performance. In addition to the spectrum itself, several considerations and opportunities that related to the spectrum levels emerged from the design sessions, and may benefit those seeking to encourage audience participation in live music performances, including *Propriety of Interactivity,* an *Ecosystem of Extras,* and facilitating *Remote Participation*.

*Propriety of Interactivity.* Although many possibilities for interaction were identified, these interactions were carefully considered within a performance's form and narrative. Throughout the design sessions, groups described how some music performances are not meant for interaction. Thus, it is vital to consider what type of interactivity is appropriate for the size, interests, and capabilities of the audience.

*Gamification.* As with many arenas within children's technologies [4], designs throughout the three sessions considered how game elements could lead to more engaging experiences for audiences during live music performances. Turn-taking rules, rewards, punishments, and recognition dynamics could all influence audience members' ability to "level up" and contribute more to the performances.

*Ecosystem of Extras.* Related to the gamification of live music performances, many designs saw the incorporation of "extras" such as clothing, food, art, and prizes to integrate audiences fully into the performance. Extras were not solely included for interaction; for instance, in the first session, a group saw value in "dressing up" for the show, a parallel to a performer putting on a costume. Similarly, groups saw value in crafting a community artifact as part of or as a result of the performance to enhance audience participation.

*Remote Participation.* In the interest of making live performances available and interesting to persons who are not able to physically attend, different interaction opportunities were designed for remote audience members. Remote audience members were included in both active (e.g., direct digital feedback) and passive (e.g., heart rate monitor) methods of audience participation in the performance.

### Limitations and Future Work

This exploratory study was limited to three, 90-minute design sessions with 6-8 child designers, and therefore future work will be required to validate and expand upon the preliminary findings presented here. Future work will consider how components of the spectrum may be integrated into traditional and new music mediums for children and adults.

## Acknowledgements

We would like to thank the adult and child design partners of the University of Maryland's Kidsteam.